# Use of Bayesian Network characteristics to link project management maturity and risk of project overcost.

Felipe Sanchez
*Sopra Steria*
Paris, France
felipe.sanchez@soprasteria.com

Davy Monticolo
*Université de Lorraine, ERPI, EA 3767*
Nancy, France
davy.monticolo@univ-lorraine.fr

Eric Bonjour
*Université de Lorraine, ERPI, EA 3767*
Nancy, France
eric.bonjour@univ-lorraine.fr

Jean-Pierre Micaëlli
IAE Lyon School of Management,
Lyon, France
jean-pierre.micaelli@univ-lyon3.fr

*Abstract*—The project management field has the imperative to increase the project probability of success. Experts have developed several project management maturity models to assets and improve the project outcome. However, the current literature lacks of models allowing correlating the measured maturity and the expected probability of success. This paper uses the characteristics of Bayesian networks to formalize experts' knowledge and to extract knowledge from a project overcost database. It develops a method to estimate the impact of project management maturity on the risk of project overcost. A general framework is presented. An industrial case is used to illustrate the application of the method.

I. INTRODUCTION (HEADING 1)

Project management maturity models (PMMM) are considered as a framework for assessing organization's project management competences, then designing the planning of actions that may result in an improvement of these competences, thus, in a better project performance. Literature exposes several examples of companies apply PMMM to their business [1].

PMMM were conceived with higher expectations. Academics and practitioners couldn't identify that more project maturity would lead to better project performance [2], [3]. Even if there is some evidence that higher-level maturity positive correlates to certain success criteria (e.g. market share) [4], [5], there is no empirical evidence that the level of maturity affects significantly the project success [6]. Several studies could not identify significant correlation between the maturity of the project and the project success [7], [8], and they cannot find empirical justification on project maturity contribution to project performance or long term organizational success [9]. Lastly, literature emphasizes the lack of enough evidence in growing project management maturity with increasing the project management success [10].

To fill this gap, we propose an integration process using a general maturity evaluation framework, grounded in the common theoretical framework of PMMM and Bayesian Networks.

Project management data are usually scarce and incomplete, and making good decision from previous data is a main overall challenge of this research. While classical machine learning techniques like neural networks give answer based on available data, Bayesian network include non-sample or prior human expertise that is relevant. Usually, an interview with an expert asking for the impact of several parameters is the best manner to recollect all that information that makes the them richer than the other techniques. Combining this expertise with sample data produce a powerful technique that can generate a realizable and performant enough model.

Bayesian networks are recognized as a powerful tool for risk analysis and decision support in real-world problems. Their diagnosis, prediction and simulation consist of observational inference of conditional probability relations. Bayes networks has been used to predict failure in complex systems specially medical systems, also the ability to explicitly model uncertainty makes them suitable to an enormous number of applications in a wide range of real-world problems including risk assessment [11], bankruptcy prediction [12], product acceptability [13] or medicine [14]. Project Management can be conceived as a complex system suffering from several diseases that can be diagnosed and avoided by working on the best project management practices in the right context.

This research will focus on answering these questions: How it is possible to use previous literature on PM maturity and performance relationship to build a more inclusive model? And, how it is possible to use correlation coefficients to build a Bayesian networks?

Consequently, this paper aims to present a method for using project management maturity models to predict project overcost. An example is presented with a study case related to oil and gas projects to illustrate the possible uses.

The paper is organized as follows: in Section II we present the theories related to project management maturity levels and Bayesian networks. Section III describes the method proposed and the framework used to evaluate project management. Section IV presents a case study. Finally, in section V exposes conclusions and perspectives for future research are discussed.

## II. RELATED WORK

### A. Maturity Evaluation

The concept of project management maturity was born from to the concept of process maturity [15]. The Software Engineering Institute of Carnegie-Mellon University was a pioneer in implementing one, called the Capability Maturity Model (CMM) [16]. These models should be an indicator or measurement of the ability of an organization dealing with projects [17], [18]. In order to achieve the right completion of the project, the maturity model requires a benchmarking of practices in several maturity levels.

Most of the model describe maturity in a perfection scale structure of five maturity levels. The first step represents informal project management, while the upper levers represent implemented and improving project management processes. The assessment of project management organization is usually done by checking whether the actors accomplish best practice [15], [19], [20] based on a defined Project management body of knowledge, usually the PmBOK from PMI.

There is a consensus among the researchers in using the PMBOK terms and theoretical constructions [21]. For instance, Project Management propose 6 Process Groups: Initialing, Planning, Executing, Monitoring and Controlling, Closing. PMI also proposes 10 knowledge areas. This model consists in a product that should be generated by assessing each one of the 49 project management process and their sub processes. [22]

### B. Bayesian Networks Review

Literature had made correlations between the same project management factors and project performance. Techniques include bi-variate correlation and multiple regression tests [24], structural equation modeling [25], [26], or Bayesian Networks (BN) [23]. The present study will focus on BN.

A Bayesian Networks (BN) is a type of probabilistic graphical model that represents a set of variables and their dependencies described by a graph. A graph is a pair G = (V, E), where V is a finite set of distinct vertices and E ⊆ V×V is a set of edges. An ordered pair (u, v) ∈ E denotes a directed edge from vertex u to vertex v, and u is said to be a parent of v and v a child of u [27]. BN are direct acyclic graphs (DAG) representing probabilistic relations between variables in which the nodes represent variables and the arcs express the dependencies between variables. Where the nodes represent the states of random variables and the arcs pointing from a parent node to a child node represents the causal condition dependency between two nodes. This relationship is represented by the probability of the node's state provided different probabilities of the parent node's state.

Bayes Networks have the particularity of using conditional probability to describe events. Any belief about uncertainty of some event or hypothesis H is assumed provisional. This is called *prior probability* or 'P(H)'. This prior probability is updated by the new experience '*e*' providing a revised belief about the uncertainty of 'H'. The new probability is called *posterior probability* given the evidence or 'P(H|e)'. The Bayes theorem (1) describes the relationship of posterior probability and the prior probability.

Where 'P(*e*|H)' is the probability of the evidence been true given that the hypothesis is true, and 'P(*e*)' represents how probable is the new evidence under all possible hypotheses.

$$P(H|e) = \frac{P(e|H)*P(H)}{P(e)} \quad (1)$$

The conditional probability table (CPT) exposed the probabilistic relationship between nodes. Figure 2 illustrate this concept on the three basic directed acyclic graph (DAG) models [28]. In this example, each node has two states True (T) and F (False). This figure also exposes the different CPT for each nodes' configuration.

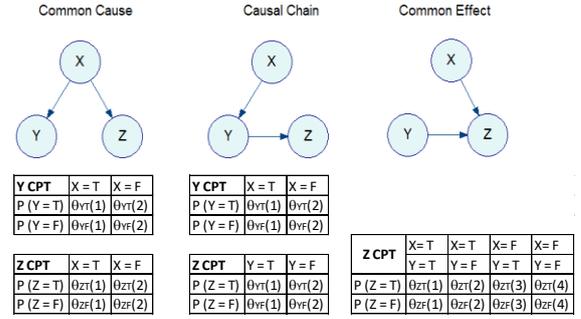

Fig. 1. The three basic DAG models

For a BN the conditional probability distribution can be represented as a set of parameters $\theta_{ik}$, where $i = 1,..,n$ defines each variable; $k = 1,...r_i$ defines each of the $r_i$ variable values; $j = 1,…q_i$, defines the set of $q_i$ variable configuration of parent variable values. For each of the $q_i$ valid configurations of parent variable values of node $i$, there is a multinomial distribution with vector parameter $\theta_i(j)=\{\theta_{i1}(j), \theta_{i2}(j), … , \theta_{ir}(j)\}$. For instance, in the *Common Effect* graph of figure 1, parameter $\theta_{ZT}(2) \equiv P(Z=T \mid X=T, Y=F)$, represents the conditional probability of Z to be true given that X is True and F is False.

Full parameterization of the BN requires the complete estimation of the vector parameter $\theta$. Subsequently, two types of analysis are possible: backwards inference which allows given an observation to fin the most probable cause among the hypothesis (diagnosis), and top down inference which allows to estimate the probability of an observation given the assumptions (prediction) [13].

### C. Uses of Bayesian Networks in Project Management evaluation

An outstanding research uses Bayesian networks for evaluating projects' overruns [23]. Its data comes from a survey of project managers in several companies. The author built a network that has six variables as input and exposes four variables as output. All nodes have two states: True/False, Yes/No. This model uses two states for each of its 26 nodes. It has six input nodes: Use of technology, strict quality requirements, multiple contracts, multiple stakeholders and variety of perspectives, political instability and susceptibility to natural disasters. It has four output nodes: Decrease in quality, low market share, time overrun and cost overrun. This model explore the interdependency between project complexity features, risk and project objectives, but it doesn't show any perfection scale (maturity level) capable of taking measures from the task of the project. Still, in this network there is not a measure of observable variables, i.e. project practices.

In addition, we found an example of where requirement definition maturity levels are correlated with project management succeed metrics [29]. It defined a correlation between requirements maturity and different metrics for project performance. This study let us compare the effect of different states of requirements maturity in selected criteria of project performance. The network has one input: Requirements Maturity and five direct outputs. It shows the probability of each type of overruns based on one measure. However, the network does not show how this maturity has been measured, or which criteria are needed to reach each level.

We show two examples of how BN as a modeling tool might manifest a better causal map of interrelationships in project management literature. In the next session, we will be building a new evaluation framework based on common characteristics of project management.

## III HOW TO STRUCTURE BAYESIAN NETWORKS TO LINK PMM AND PROJECT PERFORMANCE

Most of the components of the method are based on establish work, its novelty provides an aggregation framework of BN characteristics that has never been used in the area of project management maturity evaluation for performance correlation. Project performance can be measured by several kind of indicators i.e. price, quality, delay [30], [31]. In this research we used project overcost.

The purpose of the current investigation is the development of a methodology capable of correlating maturity with process performance. As exposed before, linking project management maturity with performance is a highly complex problem. However, our first approach, to solve this problem is by proposing the use of intermediate nodes, here called "*drift factors*", that would link project management maturity levels with the probability of low performance. This session will explain in detail how our methodology is defined.

We have difficulties in finding a structured database of project management maturity evaluation. The companies usually have database of less than hundred projects. For a learning algorithm, this means a low quantity of training examples. The proposed methodology is specially adapted when the quantity of data is small because it uses databases as well as expert knowledge.

First, it will select criteria that may produce economic drift in projects. Then, it will link those criteria to PMMMs, proposing this hypothesis: the more mature the project is, the less probable it is to fall into a drift. Finally, this research will create a Bayesian network to simulate and test the hypothesis using data from engineering projects. The simulation results can assist the decision-making process by testing different project management strategies by measuring their maturity and their impact on projects performance.

### A. Define an invariant framework for project management evaluation

An invariant-based maturity model is a normalized framework. In this structure, the project process groups can be divided in three delimited time structures, called chronologies. Hence, for each domain it is defined the time characteristics based in the phase of the project: Prepare Monitor and Valuate. Otherwise, activities and routines inside the chronologies have common characteristics, such as they are doing by someone (resource invariant), they are doing in a specific time, or they may have a repetitive nature (frequency invariant) and they have several levels of detail and granularity (activity granularity invariant). Figure 2 shows the framework for frequency invariants.

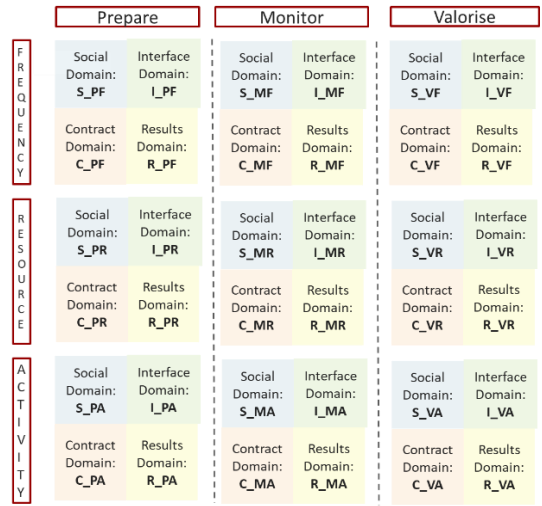

Fig. 2. Invariant-based PMMM.

### B. Define the states of the BN nodes that measure project management maturity.

For project management evaluation, we used an invariant-based PMMM framework, shown in figure 2. This framework classifies the maturity measures in states of the same nature.

As we are talking about maturity level, we would tend to associate each state of a node with each maturity level. However, this is not efficient. When learning the prior probabilities of the states from the data it is important to guarantee that there is the right number of states relative to the sample size. If the number of states was high, the TPC would increase the need for data, in that case, the model perform well on training data but it will not perform well on new data. Under these circumstances, the number of states should reduce. Typically, high-performing BN has two states in most of the variables. Else, *"If there are too many states, the model will suffer from high variability, which typically results in model overfitting."* [14]

In the proposed model, exposed partially in Figure 3 (only for Monitor-Resources invariants) each maturity level represents an independent node, each one with two sates: '*Yes*' or '*No*'. Each state corresponds to the answer of a maturity assessment question, defining whether a practice or routine is executed or not.

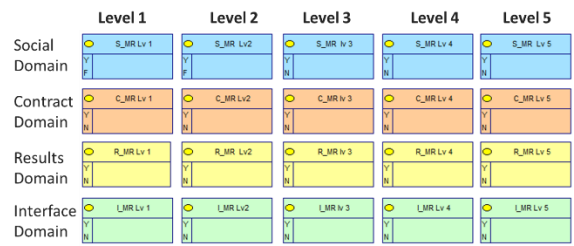

Fig. 3. Maturity level for Monitor - Resource invariants (MR) converted to nodes of a Bayesian network with two states.

## C. Define synthetic nodes (used as drift factors)

Another limitation building predictive models from project management research is also presented in the studied conducted by Mir et al. [24]. Their research let us build the network exposed in the Figure 4. Even if there is a clear target node, the main limitation of using this network comes from the underlying assumption of treating criteria as interdependent factors going from each one of the causes' nodes to project success node. Nevertheless, the model will lose its utility because the number of combinations needed to complete the CPT of 'Project Success' node is too large to be accurate, for example if 'Project Success' has two states (True/False) the vector θ would need 4096 parameters to complete its CPT table, it is has five states, it will need 244 million.

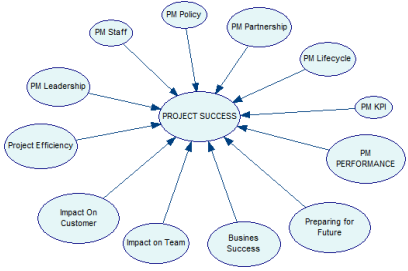

Figure 4. Research results displayed as BN [24]

Networks resulting from questionnaires and interviews, i.e. maturity evaluation, incorporate mutually exclusive variables. These variables can be modeled in a BN as the set of states of another single variable. It would be needed to define the group of the causes by a common characteristic where the causes are mutually exclusive. It is necessary to check the correlations and create states where the mutually independent variables are in the same node. However, the structure may be too complex for any combination of nodes and number of states to be able to define accurate conditional probability tables. In this case, the introduction of synthetic nodes in the network could simplify the need of data for the CPT. *"A synthetic node is one which is simply defined by the value of its parents nodes using some expert driven combinational rule."* [14]. This variety of nodes are useful for reducing the model complexity and the effects of combinatorial explosion in the CPT, and improving the structure of the model allowing the experts to visualize cause-effect relationships. As an example of using synthetic nodes, figure 5 exposes a proposition of the synthetic nodes that could simplify the network based on the study from [24]. As a result, the size and complexity of CPT in the 'PM Performance' node is reduced vastly, increasing the computation speed of and accuracy the model.

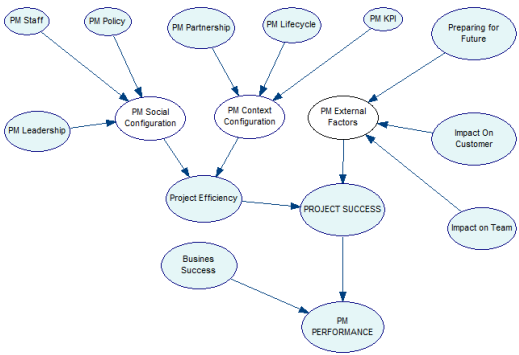

Figure 5. How the BN is transformed after introducing synthetic nodes.

In our methodology, the project management maturity levels are single nodes correlated with synthetic nodes (in this research called drift factors). Figure 6 shows how each drift factor has five parent nodes for each domain (Social, Contract, Interface, Results). This occurs for each of the nine maturity configurations: Prepare-Actions (PA), Prepare-Resources (PR), Prepare - Frequency (PF), Monitor - Actions (MA), Monitor - Resources (MR), Monitor - Frequency (MF), Valorize-Actions (VA), Valorize -Resources (VR), Valorize - Frequency (VF).

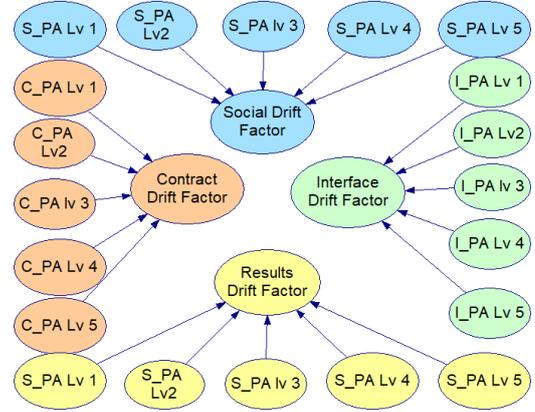

Fig. 6. Aggregation of maturity levels nodes.

## D. Define Aggregation rules and aggregation weight distribution

Summarizing, in previous steps, each drift factor node is parent of the five maturity levels nodes. The next step defines how they are correlated. We use expert knowledge to define how each maturity level is correlated with each drift factor. Table 1 is a construction extracted from the expert knowledge. It shows the aggregation weight from maturity levels to drift factors. This table shows how each maturity level contributes progressively to the probability of reaching a drift, that is, higher maturity levels implies lower drift probability.

| Maturity Level | Probability of Avoiding a Drift if the Level is Reached |
|---|---|
| LV5 | 40% |
| LV4 | 30% |
| LV3 | 15% |
| LV2 | 10% |
| LV1 | 5% |

Table 1. Maturity aggregation weights.

Once the expert choose the aggregation weighs, we can build a CPT for each drift factor node. Table 2 show this construction. Each maturity level has two states (Y/N), the combination of them goes from all N to all Y. P (Drift = T) Indicates the probability of the drift to be true, P (Drift = F) is it complement.

| PR LV1 | N | N | N | N | N | N | N | N | N | N | N | N | N | N | N | N |
|---|---|---|---|---|---|---|---|---|---|---|---|---|---|---|---|---|
| PR LV2 | N | N | N | N | N | N | N | N | Y | Y | Y | Y | Y | Y | Y | Y |
| PR LV3 | N | N | N | N | Y | Y | Y | Y | N | N | N | N | Y | Y | Y | Y |
| PR LV4 | N | N | Y | Y | N | N | Y | Y | N | N | Y | Y | N | N | Y | Y |
| PR LV5 | N | Y | N | Y | N | Y | N | Y | N | Y | N | Y | N | Y | N | Y |
| P(Drift = T) | 1 | 0.6 | 0.7 | 0.3 | 0.85 | 0.45 | 0.55 | 0.15 | 0.9 | 0.5 | 0.6 | 0.2 | 0.75 | 0.35 | 0.45 | 0.05 |
| P(Drift = F) | 0 | 0.4 | 0.3 | 0.7 | 0.15 | 0.55 | 0.45 | 0.85 | 0.1 | 0.5 | 0.4 | 0.8 | 0.25 | 0.65 | 0.55 | 0.95 |

| PR LV1 | Y | Y | Y | Y | Y | Y | Y | Y | Y | Y | Y | Y | Y | Y | Y | Y |
|---|---|---|---|---|---|---|---|---|---|---|---|---|---|---|---|---|
| PR LV2 | N | N | N | N | N | N | N | N | Y | Y | Y | Y | Y | Y | Y | Y |
| PR LV3 | N | N | N | N | Y | Y | Y | Y | N | N | N | N | Y | Y | Y | Y |
| PR LV4 | N | N | Y | Y | N | N | Y | Y | N | N | Y | Y | N | N | Y | Y |
| PR LV5 | N | Y | N | Y | N | Y | N | Y | N | Y | N | Y | N | Y | N | Y |
| P(Drift = T) | 0.95 | 0.55 | 0.65 | 0.25 | 0.8 | 0.4 | 0.5 | 0.1 | 0.85 | 0.45 | 0.55 | 0.15 | 0.7 | 0.3 | 0.4 | 0 |
| P(Drift = F) | 0.05 | 0.45 | 0.35 | 0.75 | 0.2 | 0.6 | 0.5 | 0.9 | 0.15 | 0.55 | 0.45 | 0.85 | 0.3 | 0.7 | 0.6 | 1 |

Table 2: CPT for each drift factor node.

*E. Define the network and execute parameters learning.*

Finally, it is necessary to define the joint probability distribution between the drift factors and the performance variables. Here performance is measured by the probability of over costs. This step implies the application of a machine learning algorithm e.g. naïve Bayes to the project overcost database.

## IV. CASE STUDY: EVALUATION OF DRIFT FACTORS IN OIL AND GAS OFFSHORE PROJECTS

The main purpose of presenting this case study is not to generalize a model representing a list of variables of drift and their interdependencies applicable to any project, because each project and relevant circumstance would drive the structure of the network and the weight of the variables in a different manner. We aim to demonstrate how practitioners can implement can implement the maturity modeling approach exposed previously within the context of their projects and adopt the proposed approach to prioritize tasks that improves maturity and mitigate drifts.

Therefore, in this section, we will follow the methodology to build the Bayesian network based on both expert knowledge and data from project evaluation. After the development of the initial model, numerous iterative analyses were conducted and various modifications were made accordingly to achieve a model that best suits the collected data and supports the theory.

*A. Define an invariant framework for project management evaluation*

In this case study, for project management evaluation, we used an invariant-based PMMM framework, shown in figure 3. However, we adapt this framework to the available information of our database. This reduce the number of input nodes and drift factors.

*B. Define the states of the BN nodes that measure project management maturity.*

In order to build the Bayesian network this work uses decomposition by each of the maturity evaluation framework where all Maturity levels nodes has two states, 'Yes' or 'No' respectively showing if a practice is put in place or not. Meanwhile, drift factor nodes have binary states of 'True (T)' or 'False (F)'.

We present one project performance metric, the probability of cost overrun. It has four levels of over costs: $P\_1\_$: less than 1% of total project overcost; $P\_1\_10$: between 1% and 10%; $P\_1\_10$: between 10% and 100%; and $P\_100\_$ more than 100%.

*C. Define synthetic nodes (used as drift factors)*

The current investigation involved analyzing the main causes of drift in fifteen oil offshore projects followed thru four years. The causes were selected by studying all causes of common problems and synthesizing them in the maturity system that we are proposing in this paper. 459 events were collected and then classify in the main drifts mentioned earlier. The classification was made based on the description of the event. The explored database shows the amount of money loss by each event. In order to normalize data, we calculate the percentage of losses over the cost of the project in each event, and multiply it by 100. Table 3 expose the most common drift causes in the studied projects.

| DRIFT FACTOR |
|---|
| 1.2 Late delivery from suppliers/subcontractors |
| 1.3 Late availability of ships extra costs |
| 1.4 Ship Rescheduling/Reallocation : Change of vessel |
| 2.1 Incorrect estimate of cost in tender |
| 2.2 Improper White Book Rates / Escalations |
| 2.4 Incorrect estimate of allowances/contingencies |
| 2.5 Improper Contract/Subcontract Flowdown |
| 3.1 Materials and equipment delivered out-of-specs |
| 3.2 Incomplete or partial delivery |
| 4.1 Incorrect design engineering |
| 4.2 Incorrect installation engineering |
| 4.5 Incorrect execution offshore by Acergy |
| 4.6 Incorrect execution offshore by 3rd party |
| 5. Incorrect Equipment Breakdown |

Table 3: Drift Factor for selected Oil and Gas offshore projects.

Since the evaluated project belongs to the same industry, they share specific drift causes. The model used here includes a limited number of drift factors identified by the empirical research conducted to extract the knowledge of the experts.

*D. Define Aggregation rules and aggregation weight distribution*

This initial model used a consultation with six industry experts. First, they were provided with a list of the main drift causes for these projects, to check whether the causes were meaningful and coherent with the domain they belong. Second, to ensure coherence, the experts were requested to classify the drift factor in the proposed invariant. The Figure 7 This figure show how we map drift factors and the PMMM framework, so in the final step we are able to define the structure of the BN with experts' knowledge.

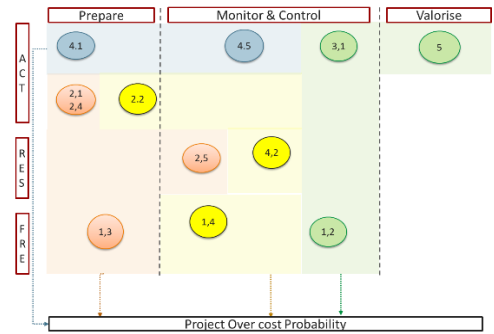

Fig. 7: Drift factor classification for oil and gas offshore projects.

*E. Define the network*

Finally, the last step is to learn the impacts of project management maturity levels on project overcosts by extracting knowledge from a database. For illustrative purposes, it is assumed that all drift factors and maturity levels converge to the node. We applied the algorithm naïve Bayes to the normalized database. This algorithm generates 16384 values corresponding to the CPT of the target node (projects over cost). The final model was created using GeNie Modeler [32]. This figure exposed the evaluated nodes It is exposed in figure 8.

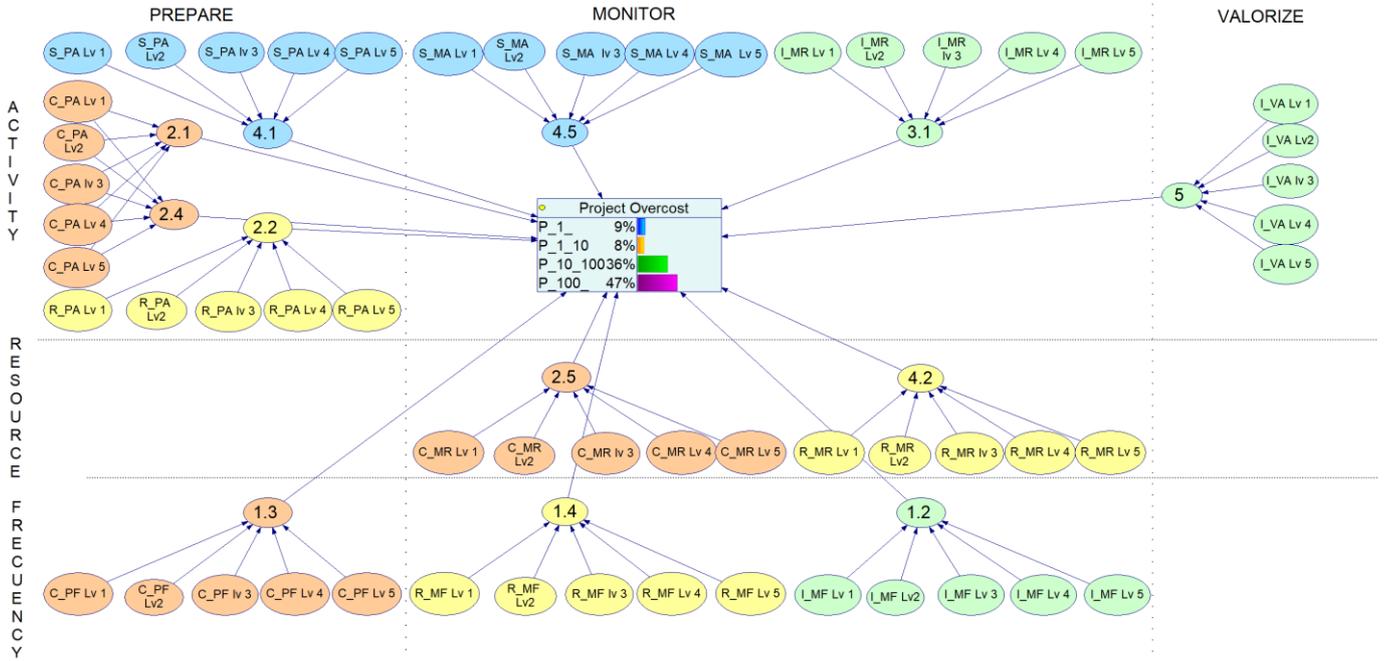

Fig. 8. Simulation model developed in GeNIe [34]

In the example, given the data collected. The project has is 9% of probability in overcosts corresponding to less than 1% of the budget. An over costs between 1% and 10% (P_1_10 in the figure) has the probability of 8%. An over cost between 10% and 100% has a probability of 36% and an over cost corresponding more than the 100% of the budget has a probability of 47%.

An additional step to exploit this BN consist in simulating actions that can be executed to improve project management maturity levels, decreasing the probability of overcost. For this, we instantiate each maturity level and drift factor to each of the level states (1 to 5) and we registered the corresponding over cost values. Figure 9 shows the results for this simulation.

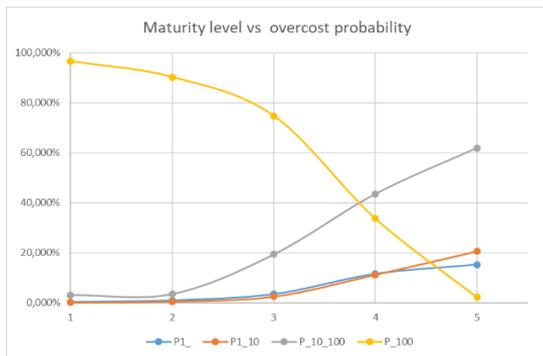

Fig. 9. Maturity levels vs over cost probability.

The simulation shows the relationship between maturity levels and percentage of the project drift. Each line represents the range of the drift and their evolution while maturity levels increase. This figure exposes how in low maturity levels (one and two) the probability of drift more than 100% (P_100) ranges between 80% and 100%. While in the upper maturity levels, between 4 and 5, the probabilities of projects are under 20%.

## V. CONCLUSION

This research aimed to develop a rigorous and repeatable method for building effective Bayesian Networks models for project management decision support from project management maturity models. In addition, it intended to exploit expert knowledge of invariant-based PMMMs to complete further data of assessment and evaluations. Furthermore, it ensure how a BN model can be used to aggregate maturity levels. Finally, we demonstrated the application of our model and the proposed modeling approach through an illustrative simulation study.

We used BN because it reveals both the variables and the path of relationship between them. This characteristic let researchers and practitioners having explicit representations. Bayes networks have advantages of others studied methods and it includes the use of prior knowledge of experts. It also reduces the risk associated with imperfect data gathering and uncertainty. BN uses an inductive mode of reasoning that permits to use both sample data and expert-judgment information in a logical and consistent manner to make inferences.

We are assuming that conditional probability values represent the capitalization of knowledge and formalization of the feedback experiences of the project management of selected projects. In this learning process, posterior knowledge of probability distribution is guided by the data of the prior information thanks to Bayesian nature of the model.

We have found that the basic underlying premise of PMMM, that higher level of maturity is, the higher the chance to complete the project successfully remains as an unproved statement. However, our simulation scenario shows that for the evaluated projects, the maturity level 4 represents lower states of probability for cost overrun. Nevertheless, further research should be done in this area.

ACKNOWLEDGEMENT

This work has been carried out under the financial support of the French National Association of Research and Technology (ANRT in French – convention CIFRE N° 2016/0778) as well as Sopra Steria. We want to acknowledge Sopra Steria Consulting and especially the "Defense and Security" Business Unit for helping us to define the proposed method.